\documentstyle[aps,prb,epsf]{revtex}  \tighten \twocolumn
\input epsf.sty

\newcommand{\beq}{\begin{eqnarray}}
\newcommand{\eeq}{\end{eqnarray}}
\newcommand{\n}{\nonumber}

\begin{document}
\draft
\title
{  
Incoherent interlayer conductivity of layered systems}

\author{ Misha Turlakov}
\address{Cavendish Laboratory, University of Cambridge, Cambridge, CB3 OHE, UK }

\address{%
\begin{minipage}[t]{6.0in}
\begin{abstract}
The criterion for the maximum incoherent conductivity $\sigma_{c,max}$
in the anisotropic system is shown to be $\sigma_{c,max} \sigma_{ab}=C (e^2/h)^2 n_{2D}$,
where $n_{2D}$ is the two-dimensional electron density, and $C$ is a numerical constant of order one.
A model of c-axis interlayer conductivity of layered system in the incoherent regime
is developed. The temperature dependence of the c-axis conductivity
is derived to be $\sigma_c (T) \sim T^\alpha$, where $\alpha$ is determined
by the effective strength of the interplane voltage fluctuations.
The ``pseudogap'' phenomenon of suppressed interplane conductivity
at low frequencies (or low voltages) is discussed.
\typeout{polish abstract}
\end{abstract}
\pacs{PACS numbers:74.25.-q, 74.25.Fy, 74.25.Gz} 
\end{minipage}}
\maketitle


The nature of the interplane transport properties of  various layered materials
(cuprates\cite{cooper,Anderson}, ruthenate $Sr_2 Ru O_4$\cite{ruthenate}, 
and some other inorganic and organic compounds\cite{others})
has been of great experimental and theoretical interest. In these systems,
the incoherent c-axis transport (defined as non-metallic temperature dependence
of resistivity and the non-Drude optical conductivity) can coexist
with metallic in-plane transport. 
For most of the cuprate compounds the c-axis conductivity is incoherent, although
for some cuprates (especially upon overdoping) the c-axis conductivity
is metallic or coherent\cite{cooper}. 
Another example is $Sr_2 Ru O_4$, 
where the incoherent c-axis conductivity is observed above $130~K$\cite{ruthenate}.

The interplane tunneling must be
viewed as a many-body process from a class of orthogonality catastrophes
when the tunneling electron unavoidably excites various degrees of
freedom of the correlated electron liquid.
In this paper I develop a model\cite{Turlakov} 
to describe the incoherent interplane tunneling
which takes into account
the effect of charge fluctuations excited during the process
of tunneling. The effect of these charge fluctuations is to decohere and
to suppress strongly the probability of interplane tunneling.

The coherent band c-axis transport (for instance, the case for $Sr_2 Ru O_4$ below $130~K$)
is well described and understood in the framework of anisotropic Drude model,
while the theory of the incoherent c-axis transport remains
controversial\cite{cooper,Anderson}. Without relying on any specific description
of the incoherent tunneling, it is possible to derive a simple estimate
for a crossover between coherent and incoherent regimes. This estimate can be made
by comparing the duration of the interplane hopping time $\tau_{hop}$
and the in-plane scattering  time $\tau_{ab}$. In the case of incoherent electron
hopping the c-axis diffusion constant $D_c$ can be written as follows
$D_c=d^2/\tau_{hop}$, where $d$ is the interplane distance.  
The Einstein relation allows
to relate the c-axis conductivity and the c-axis 
diffusion constant 

\beq
\sigma_c=\frac{e^2 \nu_{2D} D_c}{d}=\frac{e^2 \nu_{2D} d}{\tau_{hop}},
\eeq
where $\nu_{2D}=m^\star/(\pi \hbar^2)$ is the two-dimensional density of states. 
The in-plane scattering time $\tau_{ab}$ can be estimated from the standard 
Drude formula $\sigma_{ab}=(e^2 \nu_{2D} D_{ab})/d$, where
$D_{ab}=v_F^2 \tau_{ab}/2$. Thus the hopping time $\tau_{hop}$
and the in-plane scattering time $\tau_{ab}$ can be calculated
from experimental values of the corresponding conductivities
(and the density of states $\nu_{2D}$). The description of incoherent
interplane hopping is consistent if the hopping time $\tau_{hop}$
is much longer than the scattering time $\tau_{ab}$. 
It exactly means that the interplane tunneling is ``disrupted'' by the 
inelastic scattering in the planes.  
Due to the long duration of the interplane tunneling ($\tau_{hop} \gg \tau_{ab}$),
the in-plane scattering  influences 
not only the in-plane propagation but also independently 
the tunneling process. 
In other words, collective excitations
(e.g. voltage fluctuations) in a wide range of  frequencies larger than $\hbar/\tau_{hop}$ 
can be excited  by the tunneling electron. 
Note that the opposite condition is required for the validity of the coherent
hopping, namely, the in-plane scattering time $\tau_{ab}$ must be longer than 
the characteristic interplane tunneling time $\hbar/t_\perp$, where
$t_\perp$ is the tunneling matrix element. Apparently, the crossover
between two regimes happens when the interplane hopping time $\tau_{hop}$
becomes comparable or equal to the in-plane scattering time $\tau_{ab}$. 
By multiplying the left and right hand sides of the expressions for the c-axis
and in-plane conductivities given above and using the crossover condition 
$\tau_{hop} \simeq \tau_{ab}$,
we derive an estimate
for the maximum incoherent c-axis conductivity $\sigma_{c,max}$ 

\beq
\sigma_{c,max} \sigma_{ab}=C\left( \frac{e^2}{h} \right)^2 n_{2D},
\label{eq:cond}
\eeq
where $n_{2D}=k_F^2/(2\pi)$ is the two-dimensional electron density, $C$ is a constant of order 1 which
cannot be determined with the above qualitative argument. The condition  (\ref{eq:cond})
was derived from the premises of the incoherent transport. Accordingly,
it can be compared with a very similar criterion for the minimum metallic conductivity derived
from the anisotropic Drude model with simplifying assumptions\cite{Xie}.
The condition (\ref{eq:cond}) can be considered as the generalisation
of the Ioffe-Regel-Mott minimum metallic conductivity for anisotropic
layered systems. The Eqn.~\ref{eq:cond} can be rewritten as following

\beq
\sigma_{c,max} \sigma_{ab}=2\pi C \sigma_{MIR}^2,
\eeq
where $\sigma_{MIR}=\frac{e^2}{h} k_F$ is the Mott-Ioffe-Regel conductivity.
The salient point is that the crossover from the incoherent conductivity to the metallic c-axis
conductivity can happen for $\sigma_c$ significantly smaller than $\sigma_{MIR}$
(if $\sigma_{ab} \gg \sigma_{MIR}$), contrary to the naive application
of the standard Mott-Ioffe-Regel condition. The above fact that at crossover
$\sigma_c \ll \sigma_{MIR}$ was noticed widely for high-$T_c$ cuprates.

It is interesting that the condition (\ref{eq:cond}) is expressed only through
the two-dimensional (2D) density $n_{2D}$ and independent from interaction related renormalisations.
For a typical metallic 2D density $n_{2D} \sim 10^{15} cm^{-2}$ of layered materials
and choosing $C \sim 1$, the condition for the product of in-plane and
out-of-plane conductivities gives 
$\sigma_{c,max} \sigma_{ab} \sim 10^6 (\Omega cm)^{-2}$
(or equivalently, $\rho_{c,min} \rho_{ab} \sim 10^{-6} (\Omega cm)^2$).
This criterion is roughly consistent with experimental observations
of crossover between incoherent hopping  and coherent band transport.
Indeed, the crossover as a function of temperature in $Sr_2 Ru O_4$
and the crossover as a function of doping in various cuprates occurs
roughly when $\rho_{c} \rho_{ab} \sim 10^{-6} (\Omega cm)^2$.


Next I  demonstrate briefly the generalisation 
of the tunneling formalism
in the presence of the voltage fluctuations (more details can be found
in Ref.\cite{Turlakov,Devoret,thesis}). 
The voltage fluctuations which suppress strongly the tunneling
 can be related to the in-plane conductivity
$\sigma_{ab}(\omega)$ by the use of the fluctuation-dissipation theorem.
The main part of the paper is devoted to the calculation
of the specific results for the temperature and frequency dependencies
of the c-axis conductivity $\sigma_c (\omega, T)$.
At the end of the paper, I discuss experimental data which can be understood
with the help of the developed here theory.

The time-dependent tunneling formalism allows us to derive the expression
for the tunneling current $I(V)$ between two metallic planes as a function of applied voltage $V$:

\beq
I(V) & = &
\frac{2e S }{\hbar} \int dE dE' dk dk' |t_\perp (k,k')|^2 A_1(k,E) A_2(k',E')  \n \\
 & \times & \{ f(E)(1-f(E'))  P(E+eV-E',k-k')-  \n \\
 &  -     & f(E')(1-f(E))P(E'-eV-E,k-k') \}    \label{eq:big},
\eeq
where $A_{1,2}(k,E)$ are the spectral functions, $f(E)$ is a Fermi function, and
$P(E,k)$ is a Fourier transform of the tunneling probability function $P(r-r',t-t')$.
The essential new quantity here in comparison with standard treatment\cite{Mahan}
is the tunneling probability function $P(r-r',t-t')$:

\beq
&&P(\delta r=r-r',\delta t=t-t')\equiv exp(- R(\delta r,\delta t)),  
R(\delta r,\delta t) \equiv \n \\ 
&&\equiv \int \frac{d\omega d^2q}{\omega^2}
 <\delta V_{q,\omega}^2> 
(1-cos(\omega \delta t + \vec{q} \vec{\delta r})) coth\frac{\omega}{2T}, 
\label{eq:tun-prob}
\eeq
where $<\delta V_{q,\omega}^2>$ is the spectrum of the interplane voltage fluctuations.
The tunneling probability function $P(\delta \epsilon,\delta k)$ is the probability
of loosing or gaining energy $\delta \epsilon$ and momentum $\delta k$ from 
``the electromagnetic environment''
of the voltage fluctuations during tunneling. 
Perhaps, the  most serious assumption
in the derivation of Eqn.~\ref{eq:big} is that the effects of interactions 
(e.g. voltage fluctuations) on the tunneling probability $P(\delta E, \delta k)$
and the in-plane spectral functions $A_{1,2} (k,E)$ can be considered separately.

Given the spectrum of the interplane voltage fluctuations $<\delta V_{q,\omega}^2>$,
the tunneling probability $P(\delta r,\delta t)$ 
(or a Fourier image $P(\delta \epsilon,\delta k)$) can be calculated 
from Eqn.~\ref{eq:tun-prob}.  
Under certain requirements  the tunneling probability function
$P(\delta \epsilon,\delta k)$
is independent of the change of the momentum $\delta k$
implying the diffusivity of tunneling
(the momentum of the electron parallel to the planes is not conserved
after the tunneling).
These requirements have transparent physical meaning
to ensure the diffusivity of tunneling (see Ref.\cite{Turlakov,thesis}). 
The tunneling must be ``slow''
$\tau_{hop} \gg \hbar/kT$, and the in-plane conductivity $\sigma_{2D}=\sigma_{ab} d$
cannot be much larger than the quantum conductance $\sigma_Q \equiv e^2/\hbar$
(therefore the spreading of the electron density from the tunneling point
is also slow). 
The calculations are significantly simplified if the tunneling is diffusive.
We consider below only the case of the diffusive tunneling,
since these conditions are satisfied
(see Ref.\cite{Turlakov,thesis}) for the most interesting case of cuprates . 
If any of these conditions (still in the regime of incoherent hopping
$\tau_{hop} > \tau_{ab}$) is violated,
the full structure of the probability function $P(\delta \epsilon,\delta k)$
is necessary to consider to account for a partial momentum conservation.



Under the conditions of the diffusive tunneling, Eqn.~\ref{eq:big} can be simplified
by the integration over the in-plane momenta $(k,k')$:

\beq
\sigma_c(V)=\frac{e t_\perp^2 d \nu_{2D}^2}{\hbar }\frac{1-e^{-\beta eV}}{V}
\int_{-\infty}^{+\infty} d\epsilon \frac{\epsilon P(eV-\epsilon)}
{1-e^{-\beta \epsilon}}. \label{eq:last}
\eeq
In the derivation of the Equation~\ref{eq:last} we assumed additionally that:
(a) the density of states $\nu_{2D} (E)=\nu_{2D}$ is constant close to the Fermi surface
(b) the tunneling matrix element $t_\perp$ is given by the averaged value around the Fermi surface
(under conditions of the diffusive tunneling, the angular variation of
the local tunneling matrix element\cite{xiang} is not particularly important).

To model the interplane voltage fluctuations spectrum $<\delta V_{q,\omega}^2>$
we consider a pair of planes coupled by Coulomb interaction.
It can be shown that the necessary generalisation to a stack of planes
coupled by Coulomb interaction does not change the essential results.
The voltage fluctuations which most effectively ``detune'' (or decohere)
the interplane tunneling are antisymmetric with respect to the planes.
When electron tunnels between planes, these antisymmetric fluctuations are readily
excited and dampen the tunneling.
The voltage fluctuations (Johnson-Nyquist noise) can be 
calculated by using  the fluctuation-dissipation theorem
which relates them  with the frequency dependent in-plane conductivity
$\sigma_{2D} (\omega)=\sigma_{ab} (\omega) d$\cite{thesis}:
$<\delta V_{q,\omega}^2>=\frac{\hbar e^4}{\pi}V_{a}^2(q) Im \chi^a_{q,\omega}$,
 
\beq
\chi^a_{q,\omega} \simeq \frac{\sigma_{2D}(\omega)q^2}{-i\omega+\sigma_{2D}(\omega)V_a(q) q^2},
\label{eq:2d-Coulomb}
\eeq
where $\chi^a_{q,\omega}$  and 
$V_{a}(q)=2\pi e^2 (1-e^{-qd})/q$ are the charge susceptibility and
 Coulomb interaction in the antisymmetric channel in two dimensions.
%
The integration over
momentum gives the local voltage fluctuations

\beq
<\delta V_\omega^2> \simeq \frac{\sigma_Q}{\pi^2 Re \sigma_{2D}(\omega)} \omega
ln \left( \frac{2\pi Re\sigma_{2D}(\omega)q_c^2 d}{\omega}  \right),
\label{eq:local}
\eeq
where $\sigma_Q=e^2/\hbar$ is the quantum conductance\cite{units}, and $q_c$ 
is an upper momentum cutoff which we can take as the inverse screening
length $\kappa=2\pi e^2\nu_{2D}$ or the inverse of the interplane distance $1/d$.
%




Under the assumptions of the diffusive tunneling it is sufficient to calculate
the local tunneling probability $P(\delta r=0,\delta t)$, because the tunneling
probability is strongly peaked at $\delta r=0$. The function $R(\delta r=0,\delta t)$
in the exponent of the $P(\delta r=0,\delta t)$ is
$R(\delta r=0,\delta t) \simeq \frac{\delta t}{\tau_\phi}+
\alpha ln( 2\pi \sigma_{2D} \kappa^2 d/(kT) )$, where

\beq 
\alpha=\frac{\sigma_Q}{\pi^2} \int^{\omega_c}_{kT} \frac{d\omega}{\omega \sigma_{2D}(\omega)},
\label{eq:alpha}
\eeq
and $\tau_\phi=\pi \sigma_{2D}/(2\sigma_Q kT)$. 
The first term in the expression for $R(\delta r=0,\delta t)$ is determined by
the low frequency fluctuations ($\omega < kT$), while the second term is due to
the higher frequency fluctuations  ($\omega > kT$).
The high-frequency cutoff $\omega_c$ can be taken as either
the inverse of the underbarrier traversal time $\hbar/\tau_{tr} \sim 1eV$
or the frequency of the interband transitions ($\sim 1-2eV$) above which 
the c-axis conductivity increases substantially.
The tunneling probability  $P(\delta r=0,\delta t)$ allows to calculate
the temperature dependence of the c-axis conductivity from Eqn.~\ref{eq:last}.
If the low frequency in-plane conductivity is not strongly temperature independent
$\sigma_{ab} (\omega \rightarrow 0)=const$ (e.g. the residual conductivity),
the essential temperature dependence of the c-axis conductivity is

\beq
\sigma_c (T) \sim T^\alpha, 
\eeq
where 
$\alpha$ (defined by Eqn.~\ref{eq:alpha})
is very weakly temperature dependent exponent of the integrated weight of the voltage fluctuations.
The above result for the temperature dependence of the c-axis conductivity
as well as the frequency dependence discussed below are the main results of the paper.
It is important to notice that the main contributions to the exponent $\alpha$
(as can be clearly seen from the Eqn.~\ref{eq:alpha}) come
not only from small frequencies but also from high frequencies where the optical in-plane
conductivity $\sigma_{2D}(\omega)$ is small.

The results for the temperature dependence of the c-axis conductivity depend
sensitively on the low frequency fluctuations spectrum (or in other words,
on the temperature dependence
of the in-plane conductivity $\sigma_{ab} (\omega<kT)$).
For instance, if as observed for the optimally doped cuprates 
the in-plane conductivity has a strong temperature dependence $\sigma_{ab} \sim 1/kT$
with a negligible residual conductivity, the temperature dependence
of the interplane conductivity is modified to $\sigma_c (T) \sim T^{\alpha-1}$.
Another complication is that the low frequency spectrum of interplane voltage fluctuations
are affected by the small but finite interplane conduction.
Due to the interplane conduction,
the interplane voltage fluctuations are suppressed for the frequencies below
the inverse hopping time ($\omega < 1/\tau_{hop}$).
If the condition $1/\tau_{hop} \ll kT$ is satisfied which is  the case only for the strongly
anisotropic materials, we can neglect the effects of the interplane conduction
on the fluctuations spectrum. But if the hopping frequency $1/\tau_{hop}$
approaches $kT$, then the suppressed interplane fluctuations below $\omega < 1/\tau_{hop}$
do not dampen the tunneling. The account of self-consistent corrections 
from such effects 
(together with a possibility of tunneling with partial momentum conservation)
presents a complicated problem, 
and this paper is intended to capture only the main results under the simple conditions.
For the same reason, the description of the crossover from the incoherent hopping
to the coherent band propagation appears to be a difficult problem.

It is interesting to note that the formalism using the tunneling probability function
can be used to rederive the anomalous corrections to the tunneling density of states
of the disordered film\cite{Altshuler}. The voltage fluctuations in the disordered film are given
by the expression of Eqn.~\ref{eq:2d-Coulomb} with $V_a (q)$ substituted
by Coulomb interaction in the two-dimensional plane $V_{2D}(q)=2\pi/q$.
For the disordered film with a short elastic scattering time $\tau$ the in-plane
conductivity $\sigma_{2D} (\omega)$ is temperature and frequency independent
(up to the frequency $1/\tau$). The calculation of the tunneling probability
gives

\beq
P \sim exp \left[ -\frac{\sigma_Q}{\pi^2 \sigma_{2D}} ln \left( \frac{\omega_c}{kT} \right)
ln \left( \frac{2\pi \sigma_{2D} \kappa}{kT} \right) \right].
\eeq
This nonperturbative expression for the diffusive anomaly is equivalent
to the results in the literature\cite{Altshuler}.   
The fluctuation-dissipation theorem used above 
guarantees the equivalence of the two approaches, the one based on the self-consistent
linear response of the electron liquid to the tunneling electron and the other
based on the account of the voltage fluctuations.


One of the most distinct properties of incoherent transport is the frequency
dependence of the optical conductivity. The c-axis optical conductivity $\sigma_c (\omega)$
in the incoherent regime
is generally flat (or almost frequency independent) with suppression at low frequencies
in some cases. Even qualitatively, it is very different from metallic
(Drude peak centred around $\omega=0$) or localisation responses (a peak at finite frequency
with $\sigma(\omega \rightarrow 0) \simeq 0$). The optical conductivity
can be calculated in the same framework of the inelastic tunneling formalism.
%
Under the conditions of diffusive tunneling, the functional dependence of the
conductance $\sigma_c (V)=I(V)/V$ on the applied voltage is equivalent to the frequency dependence
of the optical conductivity $\sigma_c (\omega)$\cite{Turlakov}.
The c-axis optical conductivity $\sigma_c (\omega)$ can be calculated numerically
(see Fig.~1).
The magnitude of the conductivity $\sigma_c (\omega)$ is arbitrary in Fig.~1, 
because the tunneling matrix element varies for different materials
(while there is an upper bound on the incoherent conductivity (Eqn.~\ref{eq:cond}),
there is no lower bound). 
The suppression of the c-axis conductivity
at low frequencies is due to the soft Coulomb blockade effect\cite{Devoret,Altshuler}, namely,
the tunneling is suppressed by the slow diffusive spreading of the electron from
the tunneling point. The frequency scale of the suppression of the conductivity
is essentially given by $2\pi \sigma_{2D} \kappa$.

\begin{figure}
\epsfxsize=3.4in   
\epsfbox{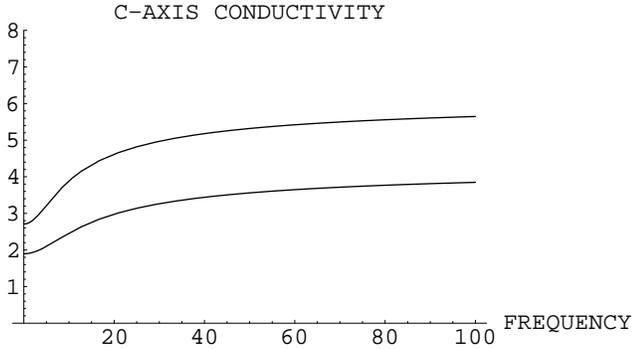}
\caption{The c-axis optical conductivity $\sigma_c (\omega)$ as a function of 
frequency $\omega$ for two different temperatures. The parameter $\alpha$ is chosen to be $1$.
The lower curve is for the lower temperature.}
\label{optic}
\end{figure}

Some systematics of the interlayer conducting properties of the cuprates can be understood
from the viewpoint of the Coulomb blocking of the tunneling by the voltage fluctuations.
The most appropriate materials (not complicated by the proximity to the regime
of the anisotropic band transport) to test the theory are the strongly anisotropic
cuprates like $Bi-2201$ and $Bi-2212$
(only inter-cell hopping which determines the transport properties is considered here). 
These materials have the incoherent
c-axis conductivity (clearly seen by the non-metallic temperature dependence
of the conductivity) throughout most of the phase diagram as a function of doping.
The effective integrated weight of the voltage fluctuations
$\alpha$ 
(Eqn.~\ref{eq:alpha})
can be determined from the in-plane optical measurements\cite{Timusk}.
Then the temperature dependence of the conductivity $\sigma_c (T) \sim T^\alpha$
can be tested experimentally.
The estimates for the parameter $\alpha$ for various cuprates give values $1 < \alpha <5$
(up to the uncertainty of the upper cutoff frequency $\omega_c$),
which are consistent with the exponents of the temperature dependence
of the c-axis conductivity.\cite{cooper}
Unfortunately, as mentioned above it is difficult to apply the theory
to some cuprates in the proximity of the crossover
to the anisotropic band propagation (as a function of the doping or temperature),
because the voltage fluctuation spectrum depends sensitively on various factors.

The pseudogap suppression of the c-axis conductivity  is explained by the proposed model
to be due to the Coulomb blocking of the tunneling.
Because with underdoping the in-plane conductivity $\sigma_{ab}$ is reduced,
the ``pseudogap suppression'' of $\sigma_c (\omega)$ is consistently expected to be more pronounced
due to the increase of the parameter $\alpha$ (in agreement with Eqn.~\ref{eq:alpha}).
Since the c-axis conductivity depends exponentially on the parameter $\alpha$,
the c-axis conductivity $\sigma_c$ is suppressed stronger than the in-plane conductivity
upon underdoping in agreement with experimentally observed increase of anisotropy
$\rho_c/\rho_{ab}=\sigma_{ab}/\sigma_c$. 
The interlayer tunneling spectroscopy experiments provide a support
for the proposed picture\cite{Krasnov}.
Other aspects of the model are consistent with optical measurements of the conductivity
$\sigma_c (\omega)$ in the normal state of cuprates\cite{Basov}:
in addition to the ``pseudogap'' suppression at low frequencies 
the c-axis conductivity $\sigma_c (\omega)$ is suppressed in a wide range of frequencies
upon lowering the temperature, and the spectral weight transferred to frequencies
above the bare tunneling matrix element frequency $t_\perp/\hbar$.

The understanding of a whole set of phenomena associated with the
pseudogap observed for the underdoped cuprates
is essential to describe consistently the c-axis transport properties
of cuprates, and this problem is not addressed here.
The applicability of the proposed theory to cuprates  is quite limited,
because the in-plane density of states $\nu_{2D}$ was assumed constant.
Nevertheless, the c-axis ``pseudogap'' suppression of the conductivity is shown to arise
due to the soft Coulomb blockade even for the constant in-plane density of states
close to the Fermi surface. Possibly, it describes the c-axis optical conductivity
of $Sr_2RuO_4$ above $130~K$
(no apparent pseudogap suppression was observed for the in-plane properties in this material). 

In conclusion, the incoherent interlayer conductivity of layered materials is proposed
to be due to the dampening of the interlayer tunneling by the voltage fluctuations.
The power-law for the temperature dependence of the c-axis conductivity
is expected, with the exponent of the power-law determined by the 
integrated weight of the voltage fluctuations.
The ``pseudogap'' suppression of the c-axis conductivity at low
frequencies (or low voltages) results from the  Coulomb blocking of the tunneling.

This work has benefited tremendously from fruitful discussions with A.J.  Leggett.




\begin{references}
\bibitem{units}
It is convenient to use throughout this paper SGS of units
in which 3D conductivity (e.g. $\sigma_{ab}=\sigma_{2D}/d$) is measured
in units of frequency ($s^{-1}$): $(\Omega cm)^{-1} \simeq 10^{12} s^{-1}$.
\end{references}
\end{document}